\documentclass[aip,jap,amsmath,amssymb,preprint]{revtex4-1}
\usepackage[utf8]{inputenc}

\usepackage{commath}
\usepackage{siunitx}

\usepackage{graphicx}
\usepackage{subfig}

\usepackage{url}

\renewcommand{\eqref}[1]{Eq.~(\ref{#1})}

\newcommand{\twopartdef}[4]
{
	\left\{
		\begin{array}{ll}
			#1 & \mbox{if } #2 \\
			#3 & \mbox{if } #4
		\end{array}
	\right.
}

\newcommand{\defeq}{\stackrel{\text{def}}{=}}

\newcommand*{\variable}{\makebox[1ex]{\textbf{$\cdot$}}}

\newcommand{\uniform}[2]{\mathcal{U}_{\left[#1, #2\right]}}

\newcommand{\ie}{\textit{i.e.}}

\begin{document}

\title{Pile-up corrections in laser-driven pulsed x-ray sources}

\author{G. Hern\'andez}
\email{guillehg@usal.es}
\author{F. Fern\'andez}

\affiliation{Departamento de F\'isica Fundamental and IUFFyM\\Universidad de Salamanca, Spain}

\date{1 December 2017}

\sisetup{range-units=single}

\begin{abstract}
A formalism for treating the pile-up produced in solid-state detectors by laser-driven pulsed x-ray sources has been developed. It allows the direct use of x-ray spectroscopy without artificially decreasing the number of counts in the detector, assuming the duration of a pulse is much shorter than the detector response time and the loss of counts from the energy window of the detector can be modeled or neglected. Experimental application shows that having a small amount of pile-up subsequently corrected improves the signal-to-noise ratio, which would be more beneficial than the strict single-hit condition usually imposed on this detectors.
\end{abstract}


\maketitle

\section*{Note}
This is a pre-print of an article published in Applied Physics B. The final authenticated version is available online at: \url{https://doi.org/10.1007/s00340-018-6982-1}. The published article includes some corrections and improvements which are missing here.

\section{Introduction}
Since the first experiments of the interaction of ultrashort and ultraintense laser pulses with matter it was realized that laser-produced plasmas may constitute a pulsed, bright, x-ray source. The alternating electromagnetic field of the laser is high enough to rapidly ionize atoms by multiphoton, tunneling or barrier suppression ionization, thus forming a plasma.~\cite{gibbon:2004} A substantial part of the plasma electrons are accelerated by several mechanisms, reaching energies up to the MeV range with a Maxwellian-like energy distribution.~\cite{batani:2008a} In their interaction with the target material x-ray pulses are produced, which spectrally consist of a continuous bremsstrahlung component and discrete characteristic line emissions. 

The x-rays energies range from few tens of keV up to several MeV, depending on the normalized laser potential and the target material. The pulse duration is of the order of hundreds of femtoseconds~\cite{radunsky:2007} and the source size can be a few times larger than the laser spot size.~\cite{batani:2008b} These properties make these sources well suited for x-ray microscopy,~\cite{giulietti:1998} phase contrast imaging,~\cite{albert:2014} spectroscopy,\cite{dorchies:2011, albert:2014} homeland security,\cite{jones:2007} and nuclear physics.~\cite{ledingham:2000, takashima:2005, albert:2014}

Since laser-produced x-rays are emitted in very short bursts, which are orders of magnitude shorter than the detector response time, experimental x-ray spectra with solid-state detectors are plagued of pile-up artifacts. This phenomenon occurs when two or more photons are detected as a single event. Thus, it represents a loss of information which may lead to a wrong analysis of the physics of the process. To avoid this problem, other types of detectors are used. Examples include include CCD detectors~\cite{seely:2011} ---although they are required work in the single-hit regime---, transmission crystal spectrometers,~\cite{seely:2011, mao:2012} indirect spectral measures with filter stacks,~\cite{chen:2009} or analyzing the Compton-scattered radiation in a material.~\cite{cipiccia:2011}

This problem is more severe when the bremsstrahlung spectrum is used to estimate the temperature of the original electron energy distribution, which can be overestimated by more than \SI{30}{\%} due to the pile-up phenomena.~\cite{zulick:2013}

A solution to the pile-up problem is to ensure the photon detection rate is sufficiently low for the probability of simultaneous photon detection in a single observation to be below an acceptable maximum value. In laser-generated x-ray emission that can only be achieved by collimating the x-ray beam or increasing the distance from the detector to the source. However, this greatly increases the observation time needed for an adequate signal-to-noise ratio.

In this work we provide a formalism to correct the pile-up in these pulsed systems ---as well as, in particular, single shot systems~\cite{cipiccia:2011}--- by using the ratio between the count rate and the repetition rate. The model requires the duration of each pulse to be much shorter than the detector response time, so all the pile-up is due to photons from the same pulse. When this condition is fulfilled the number of counts can be assumed to be given by a Poisson distribution, hence, we will call it a Poisson pile-up. The application of the method presented allows to remove the pile-up distortion in the number of counts in the channels of a detector.

\section{Theoretical methods}
\subsection{Poisson pile-up model}
Suppose a pulsed source is emitting particles with a frequency $\nu$. Assuming that the interaction probability of a particle with the detector is very small and that the number of emitted particles is much greater than one, the detection is a Poisson process and hence the number of particles producing a single count at the detector $N$ follows a Poisson distribution. Its rate parameter $\lambda$ can be estimated using the frequency $\nu$ and the count rate of the detector $r$, since
\begin{equation}
 \mathrm{E}\left[\frac{r}{\nu}\right]=\mathrm{P}(N>0)=1-e^{-\lambda}
 \,,
\end{equation}
where $\mathrm{E}\left[ \variable \right]$ denotes the expected value of a random variable.

In order to characterize the source, the detector is measuring the energy deposited in it, which follows a certain distribution $f(E)$ in the case of a single event. In general, multiple events will be recorded as a single one with an energy given by the sum of the individual energies of the detected particles. The probability density function in the case of two events is given by the convolution product
\begin{equation}\label{eq:convolution}
 (f \ast f)\left(E\right) = \int_\mathbb{R} f(x) f(E - x) \dif x
 \,.
\end{equation}
Repeated application of~\eqref{eq:convolution} describes the total energy distribution for any fixed number of independent events in an single detection. This is sometimes called the convolution power, defined by the recursion
\begin{equation}
f^{\ast n} \defeq \twopartdef{f}{n=1}{f\ast f^{\ast n-1}}{n>1}
 \,,
\end{equation}
for integer $n$.

The detected energy distribution $f_\lambda(E)$ is related with the single-event distribution $f(E)$ by applying the law of total probability to the partition of the sample space which corresponds to the outcomes of the Poisson process conditioned to $N>0$, since then no detection occurs if no particle arrives,
\begin{equation}\label{eq:poisson_series}
 f_\lambda = \frac1{1-e^{-\lambda}} \sum_{n=1}^\infty \frac{\lambda^n e^{-\lambda}}{n!}f^{\ast n}
	   = \frac1{e^{\lambda}-1} \sum_{n=1}^\infty \frac{\lambda^n f^{\ast n}}{n!}
 \,,
\end{equation}
Note that \eqref{eq:poisson_series} expresses a convex combination of distributions and hence is well defined.

To recover the undistorted distribution $f$ it is natural to apply a Fourier Transform (FT) to~\eqref{eq:poisson_series}. Denoting the FT operator as $\mathcal{F}$, applying the convolution theorem, and identifying the exponential series we find
\begin{equation}\label{eq:poisson_series_fourier}
 \mathcal{F} \{f_\lambda\} = \frac1{e^{\lambda}-1} \left( \sum_{n=0}^\infty \frac{\left(\lambda  \mathcal{F}\{f\} \right)^{n}}{n!} -1 \right)
	  =  \frac{e^{\lambda \mathcal{F}\{f\}} -1}{e^{\lambda}-1}
 \,.
\end{equation}

From this expression one can obtain either the original distribution from the piled-up one by using,
\begin{equation}\label{eq:depile_fourier}
 f = \mathcal{F}^{-1} \left\{ \frac{\ln{\left(1+(e^{\lambda} -1) \mathcal{F}\{f_\lambda\}\right)}}{\lambda} \right\}
 \,,
\end{equation}
a process we shall call depiling; as well as the distorted, piled-up distribution from the original one, 
\begin{equation}\label{eq:pile_up_fourier}
 f_\lambda =  \mathcal{F}^{-1} \left\{ \frac{e^{\lambda \mathcal{F}\{f\}} -1}{e^{\lambda}-1} \right\}
 \,.
\end{equation}

It is worth noting that, depending on the convention used to define the FT, an additional factor may appear in the application of the convolution theorem in~\eqref{eq:poisson_series_fourier}, which eventually appears in the denominator in~\eqref{eq:depile_fourier}. This fact can be simply ignored when performing calculations by imposing normalization on $f$, regardless the convention used.

Also note that $f(E)$ is the energy distribution of the events registered in the detector, which will differ from the source distribution due to the energy-dependent detector response. Typically this is corrected by pointwise multiplication with an efficiency function. Since the pile-up is a process inherent to the detection process, \ie, it depends on the interaction probability, it would be wrong to correct such an efficiency before depiling.

In order to apply the method to a realistic situation, one has to naturally replace the FT of the functions by the Discrete Fourier Transform (DFT) of the sequences, extended if needed to the origin of energies and a until a negligible high-energy tail remains if the detector window is too narrow. This increase in the number of counts has to be taken into account to estimate $\lambda$. Details on the nature of the discretization of the problem can be found in most of the theory books on the subject, cf. e.g. \S 6 in Ref.~\onlinecite{brigham:1974}. Due to the non-linearity of the transformation \eqref{eq:depile_fourier} the energy resolution of the detector decreases with the pile-up. This aspect will be discussed later in \secref{sec:resolution}. 

An additional difficulty which may arise is the presence of additive noise of a non-piled-up nature in some experimental spectra. Even in that case, the absolute value of the right hand side of \eqref{eq:depile_fourier} will provide an approximation of the depiled distribution, whose accuracy can be checked by \eqref{eq:pile_up_fourier}. Better results can be obtained by defining a sensible parametric model of the distribution ---e.g., a Maxwellian-alike model for the continuum component plus a set of Gaussians modeling the peaks--- which can be used to perform a least-squares fit using \eqref{eq:pile_up_fourier}. An example will be given below in \secref{sec:noise}.

\subsection{Analysis of noise influence}\label{sec:noise}
\begin{figure}[htb]
\centering
\subfloat[][]{\includegraphics[width=0.48\textwidth]{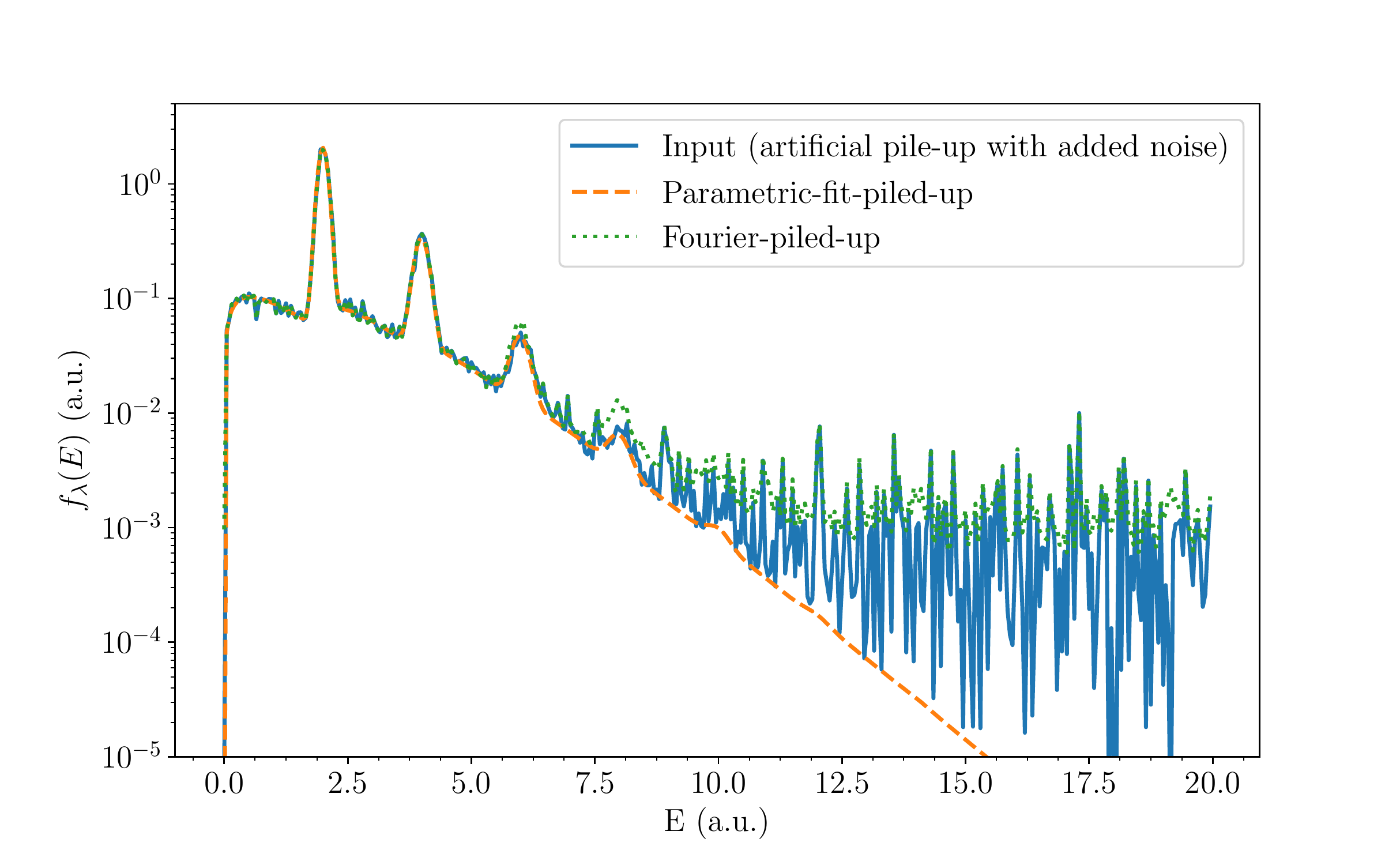}\label{fig:NoiseModelA}}\\
\subfloat[][]{\includegraphics[width=0.48\textwidth]{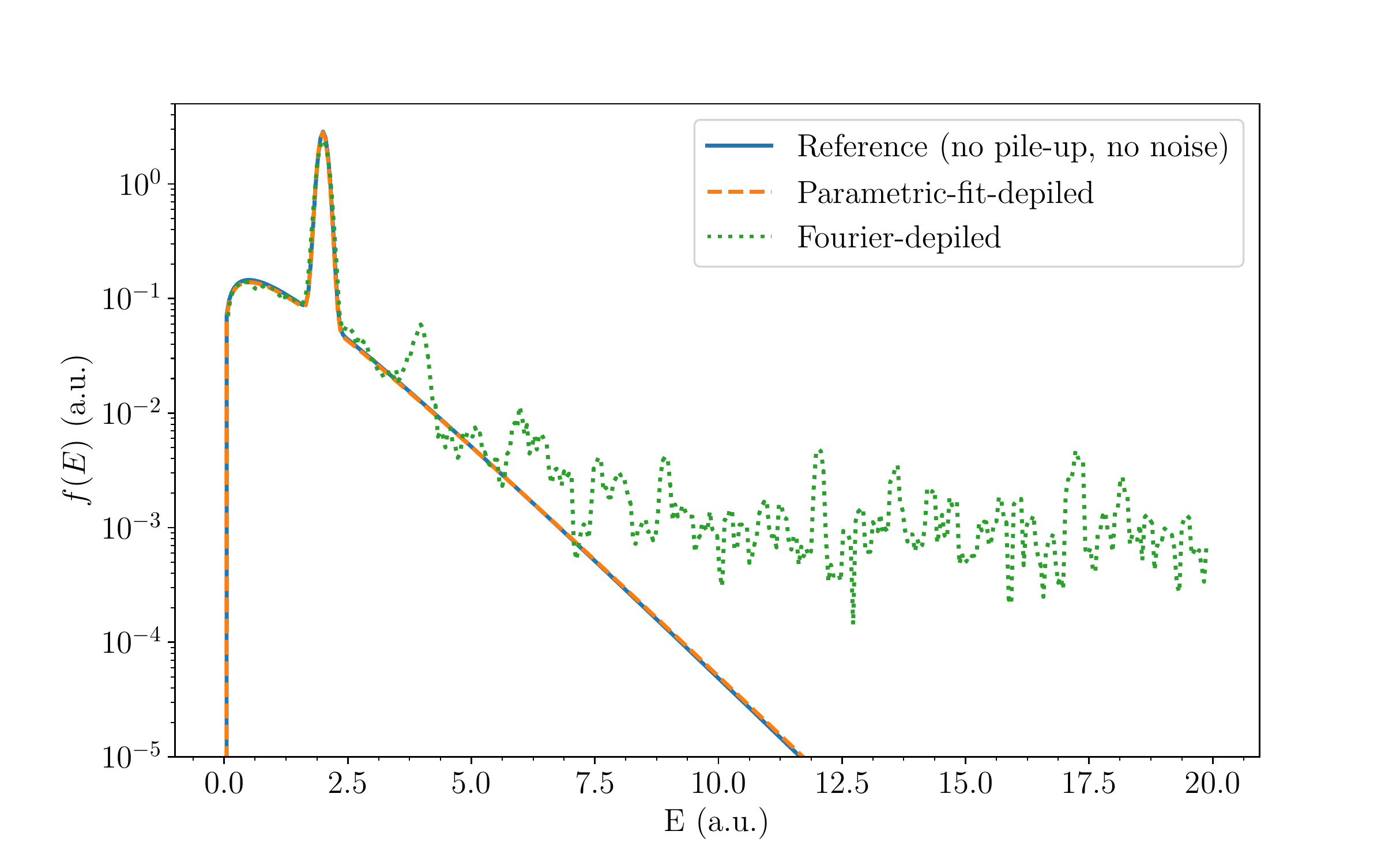}\label{fig:NoiseModelB}}
\caption{\label{fig:NoiseModel}Modelization of the impact of noise in a piled-up realistic distribution. The piled-up model and the reconstruction of the piled-up distributions $f_\lambda$, are shown in \protect\subref{fig:NoiseModelA}; while the the ideal result and the depiled distributions $f$, are shown in \protect\subref{fig:NoiseModelB}. For details on the model see the main text.}
\end{figure}

In this section we will test the behavior of the depiling process performed over a piled-up realistic distribution with some noise added to study the influence of the latter on the reproducibility of the original distribution. To account for a continuous and a discrete component, the original distribution was modeled as a convex combination of a Maxwellian, whose scale parameter is taken as the inverse of the unit of energy, and a Gaussian with parameters $(\mu,\sigma)=(2,0.1)$. This combination was sampled in bins of size 0.05, which is shown as the solid line in \figref{fig:NoiseModelB}. The distribution was piled up with a parameter $\lambda = 0.6$, chosen to qualitatively reproduce the relative heights of the peaks found in some experimental configurations. The noise was modeled by adding to each bin of the piled-up distribution a random sample of a $\chi^2_1$-distributed variable with scale parameter 0.01. The final distribution is shown as the solid line in \figref{fig:NoiseModelA}. \figref{fig:NoiseModelB} shows the direct Fourier depiling [\eqref{eq:depile_fourier}, dashed line]. Due to the added noise of non-piled-up nature, direct Fourier reconstruction neither completely removes the repetition peaks nor reproduces the asymptotic limit ---the latter being a direct consequence of such a limit being blurred by the noise in the piled-up distribution---. The best parametric fit to a general combination of a Maxwellian and a Gaussian  is also shown as the dotted line in \figref{fig:NoiseModelB}, providing a reasonable estimate of the original function, albeit the noise still influences slightly the results. Reconstructions of the pile-up spectra from both methods are also represented in \figref{fig:NoiseModelA}.

\subsection{Loss of resolution}\label{sec:resolution}

As a final note, when a detector registers a count in one channel, it is being assumed its energy is in an interval of a certain size $\Delta E$ which is called \textit{resolution}. When the pile-up phenomena occurs, this uncertainty should be considered in each of the particles that cause the pile-up event, hence increasing the difference of energies in a channel when the piled-up spectra is measured. The purpose of this section is to give an estimate on how the detector resolution decreases due the combined effect of pile-up and binning.

\begin{figure}[htb]
\includegraphics[width=0.48\textwidth]{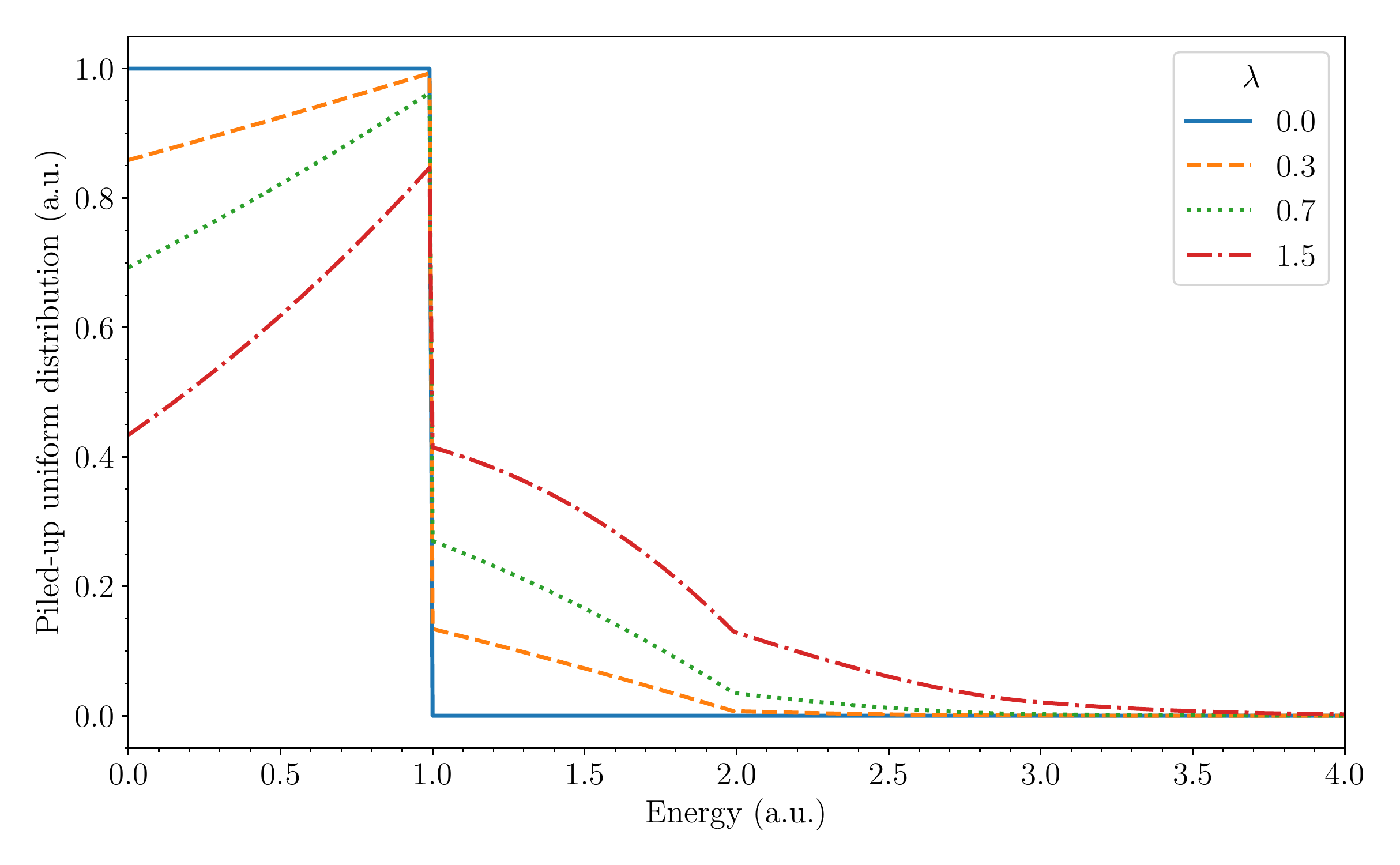}
\caption{\label{fig:UniformPileUp}Piled-up uniform distribution for some values of the Poisson parameter $\lambda$.}
\end{figure}

The exact loss of resolution depends on the (generally unknown) distribution in each of the bins. However, its magnitude can be estimated by using the piled-up distribution of a uniform distribution in a bin. Denoting the uniform distribution in $\left[a,b\right]$ as $\uniform{a}{b}$, the random variable modeling a count in channel $E_i$ is $\uniform{E_i}{E_i+\Delta E}= E_i + \Delta E \cdot \uniform{0}{1}$. The piled-up distribution deviates from sum of the lower bound energies as the pile-up of $\uniform{0}{1}$ scaled by $\Delta E$, which is a transformation of the one depicted in~\figref{fig:UniformPileUp}.

\begin{figure}[htb]
\includegraphics[width=0.48\textwidth]{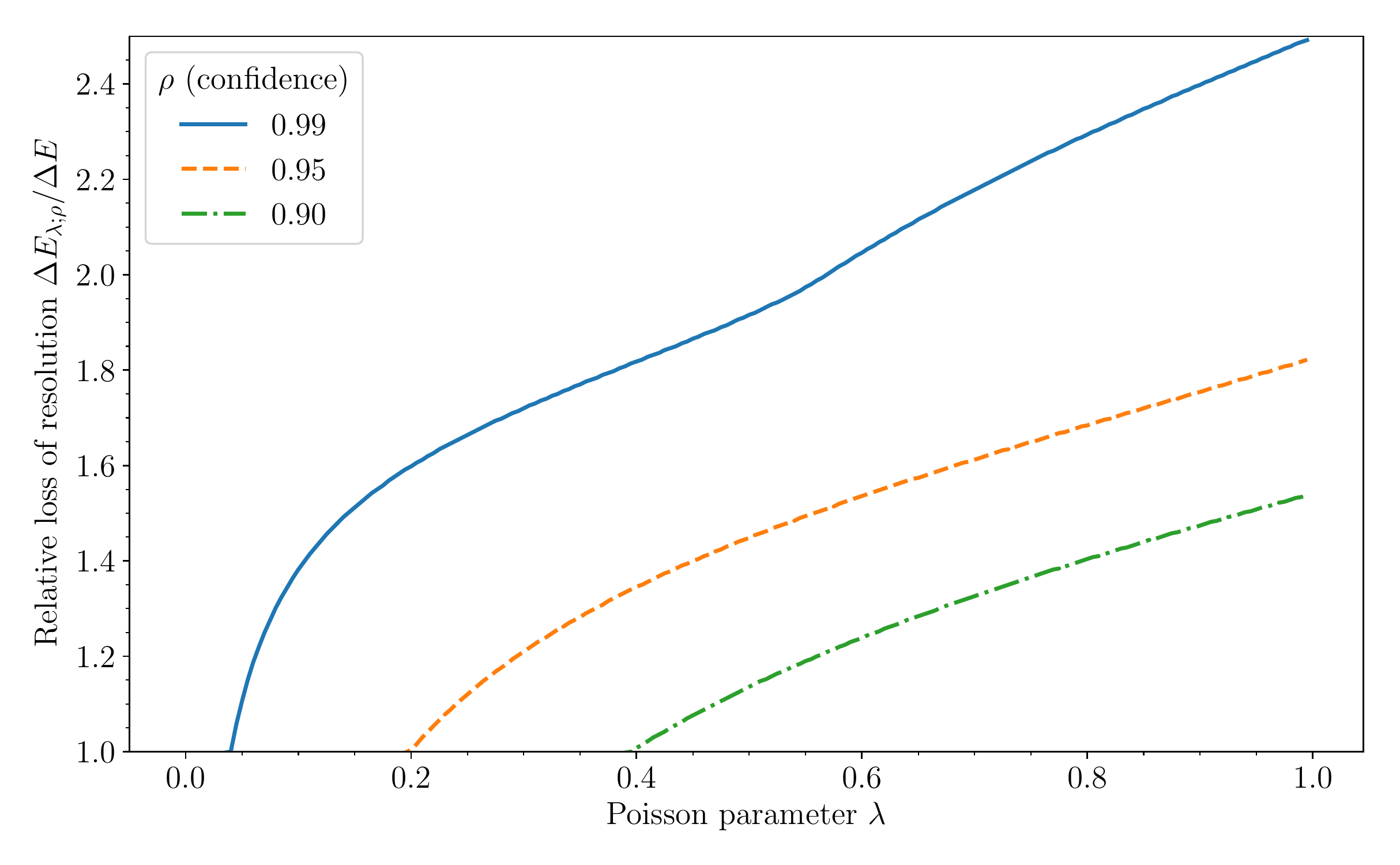}
\caption{\label{fig:ResolutionCurve}Estimation of the loss of resolution produced by combination of pile-up and binning.}
\end{figure}

The piled-up uniform distribution extends up to infinite energy, so a confidence-interval-like magnitude is needed. For that purpose we define $\Delta E_{\lambda;\rho}$ as the minimum energy range such that a fraction of counts $\rho$ from the original interval are found in it. Thus, $\Delta E_{\lambda;\rho} / \Delta E$ is the relative loss of resolution (with a `confidence' $\rho$). This is shown as a function of $\lambda$ for different values of $\rho$ in~\figref{fig:ResolutionCurve}. The curves are monotonically increasing, values below 1 indicate the change of resolution is within the energy resolution of the detector in the sense given by $\rho$. This can serve as a quantitative definition of the limit where the loss of resolution becomes non-negligible. Otherwise, the increased interval $\Delta E_{\lambda;\rho}$ can be used as an estimate of the resolution of the depiled spectra.

\section{Experimental application}
\begin{figure}[htb]
\includegraphics[width=0.48\textwidth]{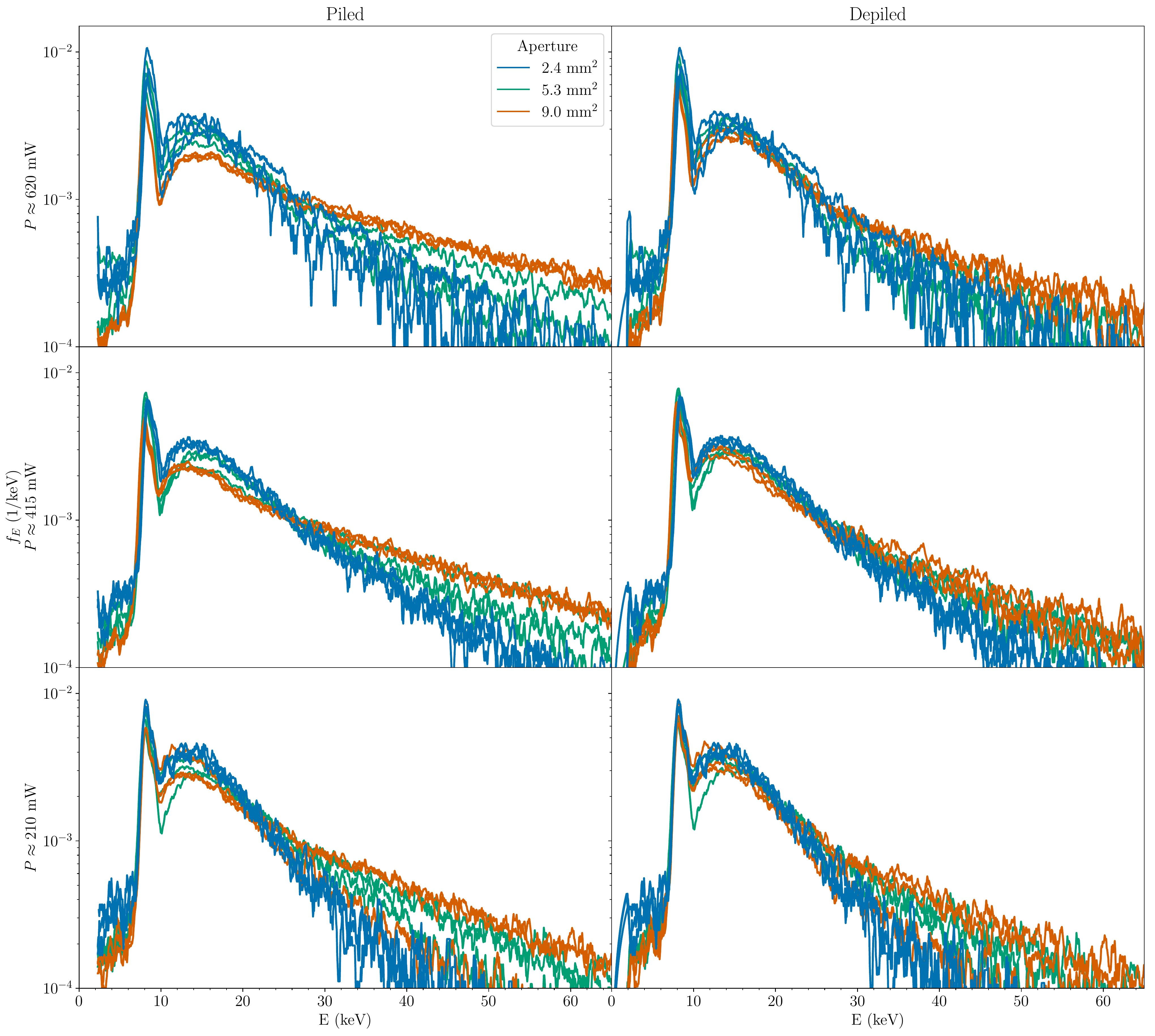}
\caption{\label{fig:ExperimentalFull}The normalized experimental spectra are shown to the left, where a different laser energy is found in each of the rows. The color of the lines indicate a different aperture, a noticeable difference is found among series. In contrast, the predictions of the depiled spectra, which are depicted to the right, are more similar to each other.}
\end{figure}

In this section the method will be applied to the experimental data obtained in the L2A2 of the Universidade de Santiago de Compostela~\cite{martin:2017} with the setup described in Ref.~\onlinecite{martin:2018}. This data was obtained using a Ti:Sa laser with an output of \SI{1}{mJ}, \SI{1}{kHz}, \SIrange{25}{100}{fs}, \SI{1E-6} contrast ratio; focused on a rotating copper target at angle of \SI{45}{\degree} towards the normal. The x-ray spectra were measured using an Amptek XR-100T-CdTe placed at \SI{19.7}{cm} from the target, in angle of \SI{16}{\degree} from the normal, against the incidence direction. A \SI{0.3}{mm} aluminum-attenuator was used to eliminate electronic low energy noise saturating the detector and to reduce the pile-up. Different series in the incidence power were made by using a half-wave plate, and three \SI{2}{mm}-thick lead diaphragms with different apertures were used to vary the amount of pile-up in each of the series. The spectra are depicted in \figref{fig:ExperimentalFull}. The Poisson parameter $\lambda$ obtained from the count rate was found to be in the \SIrange{0.01}{0.77}{} range, depending on the laser energy and the aperture of the configuration.

There are two thing worth noting in \figref{fig:ExperimentalFull}. The first is the fact that there is no evident pile-up in the aperture series until they are compared to each other. The absence of repetition peaks is not a guarantee of negligible pile-up and experimental results might still be altered by it. The other thing is the fact that the series with wider aperture show less fluctuations. This serves to illustrate that, although it is desirable to reduce the counts to reduce the pile-up, there is a compromise with the signal-to-noise ratio which can not be ignored.

\begin{figure}[htb]
\includegraphics[width=0.48\textwidth]{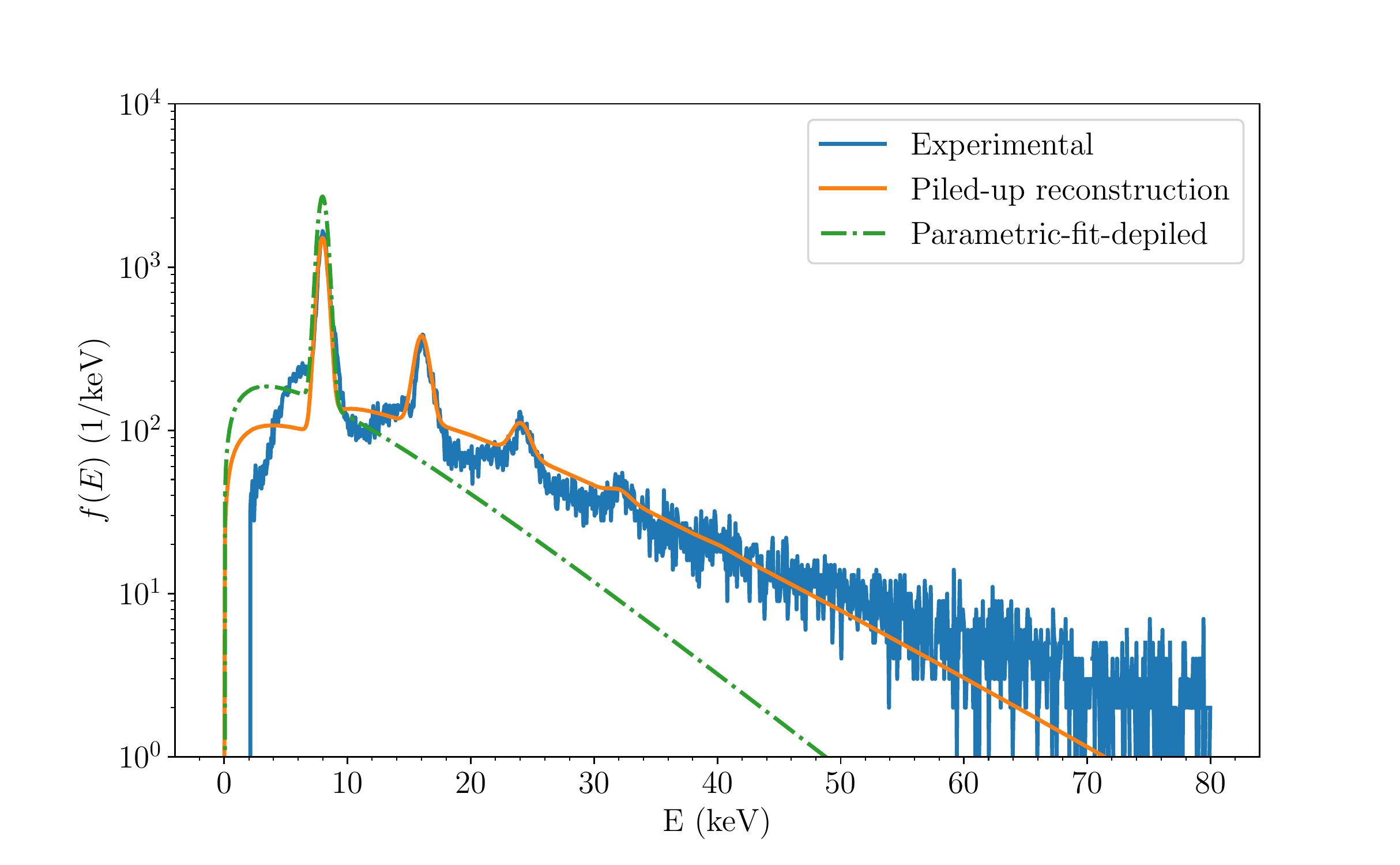}
\caption{\label{fig:ExperimentalPeaks}Reconstruction of an experimental spectra with repetition peaks using a best-parametric fit with $\lambda=1.1$.}
\end{figure}

An additional spectrum was taken replacing the aluminum attenuator by a \SI{5}{\micro m}-thick aluminum foil to allow for a higher amount of pile-up in the spectrum. This allows to analyze the model ability to resolve the repetition peaks, depicted in \figref{fig:ExperimentalPeaks}. Direct depiling with \eqref{eq:depile_fourier} fails to remove the peaks, which we think to be due to the amount of noise as discussed in \secref{sec:noise}. A parametric fit to the extremely simple functional form of Gaussian and Maxwellian is shown in the figure along with the reproduction of its pile-up. Without attempting to describe the full spectra, it is clear that the effects of pile-up like the appearance of repetition peaks and the alteration of the slope in log-linear scale ---which is typically given a temperature-related sense~\cite{key:1998}--- are reproduced.

\section{Conclusions}

In summary, a procedure for studying and resolving the pile-up in pulsed-laser driven sources has been described and verified against experimental and noisy numerical data. The results allow direct solid state spectroscopy methods to be applied without artificially decreasing the number of counts in the detector to avoid the pile-up. This method can be applied to pulsed x-ray sources such that the duration of a pulse is much shorter than the detector response time and loss of counts from the energy window of the detector can be modeled or neglected. If the count rate was unknown, the method could be still applied by adding the Poisson parameter $\lambda$ as an extra variable in a least-squares fit; however, additional evidence, e.g., repetition peak elimination, should be provided to ensure the optimal value can be attributed to pile-up and not to the uniparametric shape deformation.  An estimate of the loss of resolution due to the pile-up effect is also provided.

\section*{Acknowledgments}
The authors would like to express their gratitude to the L2A2 of the USC for providing the experimental data.
One of the authors (G. H.) gratefully acknowledges the Consejer\'ia de Educaci\'on de la Junta de Castilla y Le\'on and the European Social Fund for financial support.

\bibliographystyle{unsrt}
\bibliography{pile-up}

\end{document}